\documentclass[11pt]{article}

\usepackage{amsmath}
\usepackage{amssymb}
\usepackage{latexsym}
\usepackage{epsfig}
\usepackage{hyperref}
\usepackage{color}
\usepackage{amsthm}

\newcommand{\be}{\begin{eqnarray}}
\newcommand{\ee}{\end{eqnarray}}
\newcommand{\ba}{\begin{array}}
\newcommand{\ea}{\end{array}}
\newcommand{\bq}{\begin{equation}}
\newcommand{\eq}{\end{equation}}
\newcommand{\ft}[2]{{\frac{#1}{#2}}}
\newcommand{\ftb}[2]{{\displaystyle\frac{#1}{#2}}}
\newcommand{\bea}{\begin{eqnarray}}
\newcommand{\eea}{\end{eqnarray}}

\numberwithin{equation}{section}

\title{Noether Identities, $\beta$-functions and symmetries in DFT}
\begin{document}

\author{J. Antonio Garc\'ia$^1$ and R. Abraham S\'anchez-Isidro$^2$\\
\small Departamento de F\'isica de Altas Energ\'ias, Instituto de Ciencias Nucleares\\
\small Universidad Nacional Aut\'onoma de M\'exico,\\
\small Apartado Postal 70-543, Ciudad de M\'exico, 04510, M\'exico\\
\small $^1$ garcia@nucleares.unam.mx, $^2$ abraham.sanchez@correo.nucleares.unam.mx}

\maketitle

\begin{abstract}
Given the $\beta$ functions of the closed string sigma model up to one loop in $\alpha'$, the effective action implement the condition $\beta=0$ to preserve conformal symmetry at quantum level. One of the more powerful and striking results of string theory is that this effective action contains Einstein gravity as an emergent dynamics in space-time. We show from the $\beta$ functions and its relation with the equations of motion of the effective action,  that the differential identities \cite{callan}  are the Noether identities associated with the effective action and its gauge symmetries. From here, we reconstruct the gauge and space time symmetries of the effective action. In turn, we can show that the differential identities are the contracted Bianchi identities of the the field strength $H$ and Riemann tensor $R$.
Next, we apply the same ideas to DFT. Taking as starting point that the generalized $\beta$ functions in DFT are proportional to the equations of motion, we construct the generalized differential identities in DFT.  Relating the Noether identities with the contracted Bianchi identities of DFT, we were able to reconstruct the generalized gauge and space time symmetries. 
Finally, we recover the original $\beta$ functions, effective action, differential identities, and symmetries when we turn off the $\tilde x$ space time coordinates from DFT. 
\end{abstract}

\section{Introduction}

The momentum and winding modes of closed string on a torus suggest the idea that the effective space time description of the sigma model that include the metric $G$, the antisymmetric $B$ field and the dilaton $\Phi$,  should involve winding type coordinates as well as the usual space-time coordinates. Recently developed double field theory (DFT)  is motivated from this idea and  implements T-duality manifestly by doubling the space-time coordinates.

 T-duality acts 
on this momentum and winding modes of closed string via the non-compact duality group 
$O(d, d; Z)$. Compared to this simple action of T-duality on momentum and winding modes, 
the transformation of  the fields $G,B,\Phi$ under T-duality takes a complicated nonlinear form known as Buscher transformation \cite{Buscher}.

A deep understanding of the T-duality alter our usual intuition about geometry by relating different backgrounds with different geometric content. In the quest to understand better this fundamental property of string theory, it is quite natural to associate with momentum modes a space time coordinates $X$ and for  winding modes another space-time coordinates $\tilde X$  just defined as the canonical variables associated to the momenta and winding modes. In this double space time, the background fields are functions of all the variables $G(X,\tilde X), B(X,\tilde X), \Phi(X,\tilde X)$, but the question is if we can construct an action for these fields and its corresponding symmetries. How can we  implement the symmetries of such action? 

Recently Hull and Zwiebach \cite{Hull-Z,Hohm-Hull-Z,Hohm-Hull-Z-1} constructed double field theory from the closed string field theory based from early observations by \cite{Duff,Siegel}. In particular in the work \cite{Hohm-Hull-Z-1} the authors developed and interesting and deep formulation of DFT in terms of a generalized metric ${\cal H}$ and a generalized Riemann tensor ${\cal R}$ that have many properties analogous to the standard Riemann geometric. The resultant theory extends the geometric ideas outside of the standard formulation generalizing the gauge and space time symmetries of the space time effective action of string theory. We will call here the initial formulation based on \cite{Hohm-Hull-Z} the 
${\cal E}$-formalism and the second more geometric work \cite{Hohm-Hull-Z-1} as the ${\cal H}$-formulation. A formulation based on a frame-like construction  \cite{Siegel} was recently developed in \cite{Hohm-Kwak}  and used for a better understanding of  the ${\cal E}$ and ${\cal H}$ formalisms. This formalism is particularly efficient to construct covariant derivatives and Bianchi identities.

After these seminal works many applications and developments of DFT was worked. We will not intend to review here this active area of research but refer the reader to \cite{rev}. In the course of these developments a natural task was the construction of a double sigma model action that gives as an effective action the action of DFT in double space time  \cite{sigma-model}. A previous effort in this direction was developed in a double space-time theory \cite{Berman1} where the sigma model and $\beta$ functions was constructed. \\

Based in the ${\cal E}$ formalism,  we will recover the associated $\beta$ functions up to one loop-$\alpha'$  by just identifying the equations of motion of the DFT action with the $\beta$ functions. The relation between equations of motion and $\beta$ functions is particularly clear and  natural in DFT. Since the effective space time action in double space-time have new enhanced symmetries that implement T-duality manifestly, the symmetries of the equations of motion are also symmetries of the $\beta$ functions and in turn the $\beta$ functions are also manifestly T-duality invariant. If we change the background fields $G,B,\Phi$ using Buscher rules, we will obtain new $\beta$ functions that are of the same form as the original ones implementing the T-duality manifestly.

In particular we will show that when we turn off the added double coordinates $\tilde X$ we obtain the standard results of closed string theory effective action, recovering the standard $\beta$ functions and gauge and space time symmetries.

We also ask the question if the Noether approach to local symmetries (Noether identities) is capable of handling these new generalized symmetries that arise in the DFT framework. It is important to stress that the symmetries of DFT do not close under the standard Lie bracket but only close in the so called c-bracket that can be considered as a generalization of the standard Lie bracket. So it is not clear that the $\beta$ functions, equations of motion and generalized symmetries can be considered from the perspective of Noether second theorem. It is also important to recall that DFT is consistent if we impose the analog of the modes matching condition $L_0=\bar L_0$ of the close string sigma model but in space time. This condition in space time is called the strong condition. It is important to stress that this condition is not a consequence of the DFT action but must be imposed by hand from outside of the variational principle. The closeness of the algebra of symmetries is only possible thanks to this strong condition, so is not obvious that Noether theorems can be applied to this case. Fortunately a recent analysis \cite{Blair} (see also \cite{Hohm-Kwak,Kwak} for an approach more appropriate for our discussion)  show that in fact it is still possible to implement Noether Ideas in this context\footnote{Another covariant approach to Noether global symmetries and invariances of DFT was developed in \cite{korea1}}.

 As a first step we will investigate the relation of the $\beta$ functions
with the Noether identities using the effective space time action for the closed string sigma model. Up to our knowledge this relation between Noether identities, differential identities and Bianchi identities was not worked out previously. It is quite interesting to observe that the so called differential identities \cite{callan} are in fact the Noether identities for the gauge and space symmetries of the effective action. All what we need is to trade the $\beta$ functions by the equations of motion of the effective action and  use them into the differential identities to recover the Noether identities. From the Noether identities we can reconstruct the symmetries of the effective action.

As an application of this observation we will implement these ideas in DFT. 
In particular we will construct the generalized differential identities in DFT by showing that they are a consequence of the generalized Bianchi identities. The Bianchi identities encode the generalized symmetries of the DFT action.
To our knowledge the generalized differential identities was not considered previously.

By just defining the generalized $\beta$ functions of DFT as the equations of motion of the DFT action a consistent and very pleasant result emerge. We will show that when we remove the $\tilde X$ dependence from the action, we recover the standard formulation of the $\beta$ functions, the effective action and the gauge and space time symmetries of the usual closed string effective action. In particular from the generalized differential identities, we obtain the standard differential identities.

In section 2 we will setup our basic starting point, the relation between the differential identities, Bianchi identities and Noether identities in usual formulation of the effective action in the closed string theory. In section 3 we perform a field redefinition of the dilaton that involves the square root of the metric determinant. This redefinition allow us to write the $\beta$ functions in more compact and elegant form. In particular using this redefinition the correspondence between $\beta$ functions and equations of motion of the effective action is quite simple.  In section 4 we develop the application to DFT. 
In Appendix A we present a brief introduction to
the frame-like formalism and present a clear deduction of the generalized Bianchi identities as given in \cite{Hohm-Kwak}.

\section{$\beta$ functions and symmetry}

The $\beta$ functions of the closed string $\sigma$-model
\begin{equation}
	S=\frac{1}{4\pi}\int d^2 z \left[ \left(G_{\mu\nu}(X) +i \epsilon^{\alpha\beta}B_{\mu\nu}(X) \right)\partial_a X^\nu\partial_b X^\nu +R_{(2)}\Phi(X)\right],
	\label{modelo-sigma}
	\end{equation}
can be defined as the coefficients of the trace of the energy momentum tensor
\begin{equation}
T_a^a=-\frac12\left( g^{ab}\beta_{\mu\nu}^G+i\epsilon^{ab}\beta_{\mu\nu}^B\right)\partial_a X^\nu\partial_b X^\nu -\frac12 \beta^\Phi R_{(2)},
\end{equation}
where
	\begin{equation}
	\beta^G_{\mu\nu}=R_{\mu\nu}-\frac{1}{4}H_{\mu\nu}^2+2\nabla_\mu\nabla_\nu\Phi,
	\label{beta-g}
	\end{equation}
	\begin{equation}
	\beta^B_{\mu\nu}=-\frac{1}{2}\nabla^\rho H_{\rho\mu\nu}+
	\nabla^\rho \Phi  H_{\rho\mu\nu},
	\label{beta-b}
	\end{equation}
	\begin{equation}
	\beta^\Phi=-\frac{1}{2}\nabla^2\Phi+\big(\nabla\Phi\big)^2-\frac{1}{24}H^2.
	\label{beta-dilaton}
	\end{equation}
	Here \(H_{\mu\nu\rho}=\nabla_{[\mu}B_{\nu\rho]}\),  \(H^2_{\mu\nu}=H_{\mu\rho\sigma}H_\nu^{\rho\sigma}\) and \(H^2=H_{\mu\nu\rho}H^{\mu\nu\rho} \).
	These beta functions must be zero to implement conformal symmetry at quantum level.
	It is also true that they satisfy the so called differential identities \cite{callan}
	\begin{equation}
	\nabla^{\mu}\beta^B_{\mu\nu}-2\nabla^\mu\Phi\  \beta^B_{\mu\nu}=0,
	\label{ident-dif-gauge}
	\end{equation}
	\begin{equation}
	-2\nabla^\mu\beta^G_{\mu\nu}+4\nabla^\mu\Phi\ \beta^G_{\mu\nu}+\nabla_\nu \beta^{g}+H_{\nu\mu\rho}\beta^{B \mu\rho}-4\nabla_\nu\beta^\Phi=0.
	\label{ident-dif-difeos}
	\end{equation}
The condition $\beta=0$ can be thought as the equations of motion of the space-time fields $G,B,\Phi$. The space time action (effective action at O($\alpha'$)) whose equations of motion are equivalent to the zero $\beta$ function conditions is
\bq
S = \int dX \sqrt{-G} e^{-2 \Phi} \bigg( R + 4 (\partial \Phi)^2 - \ftb{1}{12} H^2 \bigg) \, .
\label{a-eff}
\eq
Indeed, the variation of this action is
\begin{equation}
\delta S=	-\int d X \sqrt{-G}e^{-2\Phi}\bigg(\delta G_{\mu\nu}\beta^{G\mu\nu}+\delta B_{\mu\nu}\beta^{B\mu\nu}-8(\delta\Phi-\frac14 G^{\mu\nu}\delta G_{\mu\nu})(\beta^\Phi-\frac14\beta^G)\bigg),
\label{var-a-eff}
\end{equation}
where $\beta^G$ is the trace of $\beta^{G\mu\nu}$. As a consequence, the equations of motion associated to the effective action are
\begin{equation}
\frac{\delta S}{\delta G_{\mu\nu}}= -\sqrt{-G}e^{-2\Phi}(\beta^{G\mu\nu}+2G^{\mu\nu} (\beta^\Phi-\frac14\beta^G)), \quad
\frac{\delta S}{\delta B_{\mu\nu}}= -\sqrt{-G}e^{-2\Phi}\beta^{B\mu\nu},
\end{equation}
and
\begin{equation}
\frac{\delta S}{\delta \Phi}= 8\sqrt{-G}e^{-2\Phi} (\beta^\Phi-\frac14\beta^G).
\end{equation}
We can write the $\beta$ functions in terms of the equations of motion of the effective action. The result is
\begin{equation}
	\sqrt{-G}e^{-2\Phi}\beta^G_{\mu\nu}=\frac{\delta S}{\delta G^{\mu\nu}}-\frac{1}{4}G_{\mu\nu}\frac{\delta S}{\delta\Phi},
	\label{beta-g-ec-mov}
	\end{equation}
	\begin{equation}
	\sqrt{-G}e^{-2\Phi}\beta^B_{\mu\nu}=-\frac{\delta S}{\delta B^{\mu\nu}},
	\label{beta-b-ec-mov}
	\end{equation}
	\begin{equation}
	\sqrt{-G}e^{-2\Phi}(\beta^\Phi-\frac{1}{4}\beta^{G})=\frac{1}{8}\frac{\delta S}{\delta\Phi}.
	\label{beta-dilaton-ec-mov}
	\end{equation}

The crucial observation that we want to stress here is that when these relations are used into the differential identities (\ref{ident-dif-gauge}) and (\ref{ident-dif-difeos}) we obtain the Noether identities associated with the gauge and diffeomorphism symmetries of the effective action. In fact, when we substitute the explicit relation between the $\beta$ functions in terms of the equation of motion we obtain
the Noether identities given by
\begin{equation}
	\nabla^\mu\frac{\delta S}{\delta B^{\mu\nu}}=0,
	\label{ident-noether-gauge}
	\end{equation}
	\begin{equation}
	2\nabla^\mu\frac{\delta S}{\delta G^{\mu\nu}}+H_{\nu\rho \mu}\frac{\delta S }{\delta B_{\rho\mu}}+\nabla_\nu\Phi\frac{\delta S}{\delta\Phi}=0.
	\label{ident-noether-difeos}
	\end{equation}
From here we are able to reconstruct the space-time symmetries and the gauge symmetry of the effective action
\begin{equation}
\delta_{g} B_{\mu\nu}=\nabla_{[\mu}\zeta_{\nu]}, 
\end{equation}
where $\zeta_\nu$ are the gauge parameters and
\begin{equation}
	\delta B_{\mu\nu}=\xi^\rho H_{\rho\mu\nu}, \quad \delta G_{\mu\nu}=\nabla_{(\mu}\xi_{\nu)}, 
	\end{equation}
\begin{equation}
	\delta \Phi=\xi^\mu\nabla_{\mu}\Phi,
	\end{equation}
where $\xi^\mu$ are the parameters associated with the local space-time symmetries of the effective action $S$.

Now let us analyze the content of the differential identities. Consider the differential identity (\ref{ident-dif-gauge}).
Pluggin the $\beta^B_{\mu\nu}$ in terms of the equations of motion (\ref{beta-b-ec-mov}) in this identity to obtain
\begin{equation}
	\nabla^{\mu}(-\frac{1}{2}\nabla^\lambda H_{\lambda\mu\nu}+\nabla^\lambda \Phi H_{\lambda\mu\nu})-2\nabla^\mu\Phi\  (-\frac{1}{2}\nabla^\lambda H_{\lambda\mu\nu}+\nabla^\lambda \Phi H_{\lambda\mu\nu})=0,
	\end{equation}
that imply
\begin{equation}
	\nabla^{\mu}\nabla^\lambda H_{\lambda\mu\nu}=0.
	\label{gauge-bianchi}
	\end{equation}
In turn this relation comes from the Bianchi identity of the Riemann curvature tensor, $R_{\sigma [\mu\nu\rho]} =0$. 

So we conclude that the differential identity (\ref{ident-dif-gauge}) is just the the gauge Bianchi identity associated with the $B$ field.
	


In the same way we can take the second  differential Identity (\ref{ident-dif-difeos})
and substitute the $\beta$ functions 
to obtain,
 \begin{equation}
 -2\nabla^\mu R_{\mu\nu}+\nabla_\nu R+\frac{1}{2}\nabla^\mu H^2_{\mu\nu}-\frac{1}{2}H_{\nu\mu\lambda}\nabla^\sigma H_{\sigma}^{\ \ \mu\lambda}-\frac{1}{12}\nabla_\nu H^2 =0.
 \end{equation}
The reason why this identity  is in fact true is because 
 \begin{equation}
\nabla^\mu R_{\mu\nu}-\frac{1}{2}\nabla_\nu R=0,
\label{GR-bianchi}
\end{equation}
which is a consequence of a contracted Bianchi identity of the Riemann tensor
\begin{equation}
\nabla_{[\mu} R_{\nu\rho]\sigma}^{\ \ \ \ \lambda}=0,
\end{equation}
and 
\begin{equation}
H^{\mu\lambda}_{\ \ \ \sigma}\nabla^\sigma H_{\nu\mu\lambda}-\frac{1}{6}\nabla_\nu H^2=0,
 \label{H-bianchi}
 \end{equation}
that just come from the Bianchi identity
\begin{equation}
\nabla_{[\sigma}H_{\nu\mu\lambda]}=0, \quad \to\quad H^{\sigma\mu\lambda}\nabla_{[\sigma}H_{\nu\mu\lambda]}=0 
\end{equation}

This explain in turn why the second differential identity is zero. 
 So we conclude that the differential identities (\ref{ident-dif-gauge}) and (\ref{ident-dif-difeos}) are a consequence of the Bianchi identities (\ref{gauge-bianchi}), (\ref{H-bianchi}) and (\ref{GR-bianchi}). 

This constructive argument have a nice consequence.  The symmetries associated with the effective action constraint the $\beta$ functions to satisfy the differential identities. The differential identities are the Noether identities associated with the effective action $S$. The aim of this note is to apply these ideas to DFT.

 

\subsection{ Effective action and the new dilaton}

A relevant field redefinition of the dilaton $\Phi$ in terms of a new dilaton $d$ can be used to simplify the relation between the $\beta$ functions and the equations of motion of the effective action. The redefinition is
\begin{equation}
	e^{-2{d}}=\sqrt{-G}e^{-2{\Phi}},
	\label{redef-dilaton}
	\end{equation}
and imply
\begin{equation}
\partial_\mu \Phi =  \partial_\mu d + {1\over 2} \Gamma_\mu=\nabla_\mu d\,, ~~~~\Gamma_\mu =
\Gamma^\nu_{\mu\nu} = {1\over 2} g^{\nu\sigma} \partial_{\mu}g_{\nu\sigma} \,.
\end{equation}
Notice that $d$ is a scalar density of weight 1/2.
Using this in $S$  (\ref{a-eff}) we get
\begin{equation}
\label{action-d}
S_d= \int dx e^{-2d} \Bigl[ R +  4 (\nabla d)^2   -{1\over 12} H^2\Bigr] \,.
\end{equation}
 Now we can define the new $\beta$ function after the redefinition of the dilaton as
\begin{equation}
\delta S_d=	-\int d X e^{-2d}\bigg(\delta G_{\mu\nu}\beta^{G\mu\nu}+\delta B_{\mu\nu}\beta^{B\mu\nu}+ \delta d \beta^d\bigg)
\end{equation}
where the beta functions are
\begin{equation}
	\beta^G_{\mu\nu}=R_{\mu\nu}-\frac{1}{4}H_{\mu\nu}^2+2\nabla_\mu\nabla_\nu d ,
	\label{beta1-g}
	\end{equation}
	\begin{equation}
	\beta^B_{\mu\nu}=-\frac{1}{2}\nabla^\rho H_{\rho\mu\nu}+
	H_{\rho\mu\nu}\nabla^\rho d \, ,
	\label{beta1-b}
	\end{equation}
	\begin{equation}
	\beta^d=-\frac14 ( R-4(\nabla d)^2-\frac{1}{12}H^2+4\nabla^2 d).
	\label{beta1-dilaton}
	\end{equation}
	
Notice that $\beta^G$ and $\beta^B$ are the same as the previous ones (\ref{beta-g},\ref{beta-b}) just written in terms of the new dilaton $d$, but $\beta^d$ is different. The reason behind this fact is that the $\beta^d$ function itself must be redefined by
\begin{equation}
\beta^d=\beta^\Phi-\frac14\beta^G.
\label{redef-beta}
\end{equation}
This redefinition is already implicit in the calculation of the variation of the original effective action (\ref{var-a-eff}). Indeed if we take this redefinition of $\beta^d$ in terms of $\beta^\Phi$ and the trace of $\beta^G$ and substitute the old dilaton by the new one we obtain the right result (\ref{beta1-dilaton}). The result given in (\ref{beta1-dilaton}) is just the variation of the action $S_d$ with respect to the dilaton $d$.

The differential identities (\ref{ident-dif-gauge}, \ref{ident-dif-difeos}) can be written in terms of the new dilaton  as

\begin{equation}
	\label{ident-dif-gauge-d}
	\nabla^{\mu}\beta^B_{\mu\nu}-2\nabla_\mu d \ \beta^B_{\mu\nu}=0,
	\end{equation}
	\begin{equation}
	\label{ident-dif-difeos-d}
	-2\nabla^\mu\beta^G_{\mu\nu}+4\beta^G_{\mu\nu} \nabla^\mu d+H_{\nu\mu\rho}\beta^{B \mu\rho}-4\nabla_\nu\beta^d=0,
	\end{equation}
where we have used the redefinition (\ref{redef-dilaton}) and the redefinition of the $\beta^d$ function (\ref{redef-beta}). A nice consequence of the dilaton redefinition is that  now the relation between $\beta$ functions and equations of motion is simpler 
\begin{equation}
	e^{-2d}\beta^G_{\mu\nu}=\frac{\delta S_d}{\delta G^{\mu\nu}},
	\label{beta-g-ec-mov-d}
	\end{equation}
	\begin{equation}
	e^{-2d}\beta^B_{\mu\nu}=-\frac{\delta S_d}{\delta B^{\mu\nu}},
	\label{beta-b-ec-mov-d}
	\end{equation}
	\begin{equation}
	e^{-2d}\beta^d=\frac18\frac{\delta S_d}{\delta d}.
	\label{beta-dilaton-ec-mov-d}
	\end{equation}
Pluggin these relations in the differential identities (\ref{ident-dif-gauge-d}, \ref{ident-dif-difeos-d}) we obtain the Noether identities in terms of the new dilaton 
\begin{equation}
	\nabla^\mu\frac{\delta S_d}{\delta B^{\mu\nu}}=0,
	\label{ident-noether-gauge-d}
	\end{equation}
	\begin{equation}
	2\nabla^\mu\frac{\delta S_d}{\delta G^{\mu\nu}}+H_{\nu\rho \mu}\frac{\delta S }{\delta B_{\rho\mu}}+(\partial_\nu d +\frac12\partial_\nu)\frac{\delta S_d}{\delta d}=0,
	\label{ident-noether-difeos-d}
	\end{equation}
and from here we can read the symmetries of the action $S_d$
\begin{equation}
	\label{sym-g-d}
	\delta G_{\mu\nu}=\nabla_{(\mu}\xi_{\nu)}, 
	\end{equation}
	\begin{equation}
	\label{sym-b-d}
	\delta B_{\mu\nu}=\xi^\rho H_{\rho\mu\nu}, \quad \delta_{g} B_{\mu\nu}=\nabla_{[\mu}\zeta_{\nu]}, 
	\end{equation}
\begin{equation}
\label{sym-d}
\delta d=-\frac12\partial_\mu\xi^\mu+\xi^\mu\partial_\mu d.
\end{equation}
As expected, the symmetry transformations of $G$ and $B$ are the same as the previous ones and the new dilaton $d$ transform as a density of weight 1/2.

The results of this section are straightforward but not trivial. The simple redefinition of the dilaton (\ref{redef-dilaton}) appears naturally in the formulation of the manifest T-duality invariant action in DFT as we will see in the next section.

\section{DFT ${\cal E}$-formalism}

The background-independent double field theory action can be written in terms of variable ${\cal E}_{ij} = g_{ij} + b_{ij}$ and the dilaton $d$. The action is \cite{Hohm-Hull-Z} \footnote{Here we will change notation from uppercase to lowercase letters for the metric and the antisymmetric field.}
\bq \label{action}
\ba {c}
S= \displaystyle\int dx d \tilde{x}  \ e^{-2d} \Bigg[ - \ftb{1}{4} g^{ik} g^{jl} {\cal D}^p {\cal E}_{kl} {\cal D}_p {\cal E}_{ij} + \ftb{1}{4} g^{kl} ({\cal D}^j {\cal E}_{ik} {\cal D}^i {\cal E}_{jl} + \bar{{\cal D}}^j {\cal E}_{ki} \bar{\cal D}^i {\cal E}_{lj}) \\ +  ({\cal D}^i d \  \bar{\cal D}^j {\cal E}_{ij} + \bar{\cal D}^i d \ {\cal D}^j {\cal E}_{ji} ) + 4 {\cal D}^i d \ {\cal D}_i d \ \Bigg] ,
\ea
\eq
where the ${\cal D}$ derivatives are defined as:
\bq 
\label{cali-D}
{\cal D}_i \equiv \partial_i - {\cal E}_{ik} \tilde{\partial}^k \ , \quad \bar{\cal D}_i \equiv \partial_i + {\cal E}_{ki} \tilde{\partial}^k .
\eq
To move upper and lower indices  we use  $g_{ij} = \ft{1}{2} ({\cal E}_{ij} +{\cal E}_{ji} )$.

The gauge transformations associated with the action (\ref{action}) are 
\bq \label{gauge-transf}
\ba{lcl}
\delta {\cal E}_{ij} &=& {\cal D}_i \tilde{\xi}_j - \bar{\cal D}_j \tilde{\xi}_i  + (\xi^k \partial_k + \tilde{\xi}_k \tilde{\partial}^k) {\cal E}_{ij} + {\cal D}_i {\xi}^k {\cal E}_{kj} + \bar{\cal D}_j \xi^k {\cal E}_{ik} \, , \\ [1ex] 
\delta d &=& -\ftb{1}{2}  \partial_i \xi^i +  \xi^i \partial_i d -\ftb12\tilde{\partial}^i \tilde{\xi}_i+ \tilde{\xi}_i \tilde{\partial}^i d \, .
\ea
\eq
The ${\cal E}$-symmetry can be written a more transparent form as
\be
\nonumber
\delta g_{ij}={\cal L}_\xi g_{ij}+{\cal L}_{\tilde\xi} g_{ij}+(\tilde\partial^k\xi^l-\tilde\partial^l\xi^k)(g_{ki}b_{jl}+g_{kj}b_{il}),
\ee
and
\be
\nonumber
\delta b_{ij}={\cal L}_\xi b_{ij}+{\cal L}_{\tilde\xi} b_{ij}+\partial_i\tilde\xi_j-\partial_j\tilde\xi_i+
g_{ik}(\tilde\partial^l\xi^k-\tilde\partial^k\xi^l)g_{lj}  +b_{ik}(\tilde\partial^k\xi^l-\tilde\partial^l\xi^k)b_{lj} ,
\ee
from the symmetric and antisymmetric part of ${\cal E}_{ij}$. In this form it is evident that the fundamental fields transform as expected in the $x$ space and a copy in the $\tilde x$ space plus the gauge transformations of the $b$-fields and terms that depend in $\tilde \partial$. 
 The algebra of symmetries close under a bracket called the C-bracket up to the strong constraint. This constraint take the form
$\partial^i\tilde\partial_i=0$ for all gauge parameters and products of fields or in terms of calligraphic derivatives as
\begin{equation}
{\cal D}_i A{\cal D}^i B=\bar{\cal D}_i A\bar{\cal D}^i B,
\end{equation} 
for any $A$ and $B$. For details we refer the reader to \cite{Hohm-Hull-Z}.\\

The C-bracket emerges when we try to compute the commutator of symmetries with different parameters
\be
[\hat{\cal L}_{\xi_1},\hat{\cal L}_{\xi_2}]=-\hat{\cal L}_{[\xi_1,\xi_2]_C},
\ee
where $\hat{\cal L}$ denotes the generalized Lie derivative, defined by
\be
\label{gen-lie-der}
\hat{\cal L}_\xi A_M=\xi^P\partial_P A_M+ A_P \partial_M\xi^P - A_P \partial^P\xi_M,
\ee
and $[\xi_1,\xi_2]_C$ is the $C$ bracket given by
\be
[\xi_1,\xi_2]_C^M=2\xi^N_{[1}\partial_N\xi^M_{2]}-\xi_{N[1}\partial^M\xi_{2]}^N.
\ee
We see that the last term of the generalized Lie derivative (\ref{gen-lie-der}) is unusual with respect to the standard Lie derivative acting on 1-forms. This unusual structure of the symmetry algebra give rise to our question about if these symmetries can also be handled by the Noether theorem.

The fundamental $O(D,D)$ matrix
\begin{equation}
\eta_{MN}= \begin{pmatrix} 0 &   1 \\ 1 & 0 \end{pmatrix} ,
 \end{equation}
where 1 is the D-dimensional Kronecker-delta has the property
\be
\hat{\cal L}_\xi\eta_{MN}=0,
\ee
and the capital indices are
\be
X^M=\begin{pmatrix}\tilde x_i\\ x^i\end{pmatrix},\quad \partial_M=\begin{pmatrix}\tilde\partial^i \\ \partial_i\end{pmatrix}, \quad \xi^M=\begin{pmatrix}\tilde\xi_i \\ \xi^i\end{pmatrix}.
\ee

The DFT action can also be written in the form
\be
\label{action-R}
S=\int dx d\tilde x e^{-2d}{\cal R},
\ee
where ${\cal R}$ plays the role of the scalar curvature in Riemann geometry. It is given by

  \bea\label{generalscalar}
  {\cal R}({\cal E},d) &=&
  2\left(\nabla^{i}{\cal D}_{i}d+\bar{\nabla}^{i}\bar{{\cal D}}_{i}d\right)
  +\frac{1}{2}\left(\nabla^{i}\bar{\cal D}^{j}{\cal E}_{ij} +
  \bar{\nabla}^{j}{\cal D}^{i}{\cal E}_{ij}\right)\\[0.5ex] \nonumber
  &&+\frac{1}{4}  g^{ij}
 \left(
  {\cal D}^{k}{\cal E}_{lj}\,{\cal D}^{l}{\cal E}_{ki}
  + \bar{{\cal D}}^{k}{\cal E}_{jl}\,\bar{{\cal D}}^{l}{\cal E}_{ik}
  \right)
  -\frac{1}{4}  g^{ij}\left(
  {\cal D}^{l}{\cal E}_{lj}\,{\cal D}^{k}{\cal E}_{ki}
  + \bar{{\cal D}}^{l}{\cal E}_{jl}\,\bar{{\cal D}}^{k}{\cal E}_{ik}
  \right) \\[1.0ex] \nonumber
  &&
  -\frac{1}{4}
  g^{ik}g^{jl}{\cal D}^{k}{\cal E}_{ij}\,{\cal D}_{k}{\cal E}_{kl}
  - \left({\cal D}^{i}{\cal E}_{ij}\, \bar{\cal D}^{j}d
+  \bar{\cal D}^{j}{\cal E}_{ij}\,{\cal D}^{i}d
  \right)
  -4{\cal D}^{i}d\,{\cal D}_{i}d\;,
 \eea
 and differs from the 
Lagrangian in the action (\ref{action}) by a total derivative\footnote{Given the strong constraint, one can always use an $O(D,D)$ transformation to 
  rotate into a frame where fields have no $\tilde x$ dependence,  allowing
us to set $\tilde{\partial}=0$.  In this frame we have verified that the two scalars differ by a total derivative.  We now state a simple but useful lemma.  Given two $O(D,D)$ scalars
$A (x, \tilde x)$ and $B(x, \tilde x)$ that differ by   total derivative terms,
then after an $O(D,D)$ transformation 
 they again differ by  total derivative terms.
The lemma implies
 that  the 
 original Lagrangian (\ref{action}) and  $e^{-2d} {\cal R}$ 
  differed by a total derivative before the $O(D,D)$ transformation.}


The variation of the action (\ref{action-R}) is given by
\bq
\delta S = \delta_{\cal E} S + \delta_d S \ = \displaystyle\int dx d \tilde{x}  \ e^{-2d} \delta {\cal E}_{ij} {\cal K}^{ij} \ + \ \displaystyle\int dx d \tilde{x}  \ e^{-2d} ( -2 \delta d) {\cal R} 
\end{equation}
where equations of motion are \cite{Kwak} 
\be
\label{eqm-R}
{\cal R}({\cal E},d)=0,
\ee
and
\bq \label{eqm-K}
\ba{lcl}
{\cal K}_{ij} &=& \ftb{1}{4}  (\nabla^k {\cal D}_k {\cal E}_{ij}  + \bar{\nabla}^k \bar{\cal D}_k {\cal E}_{ij} - 2 \nabla ^k {\cal D}_i {\cal E}_{kj} - 2 \bar{\nabla}^k \bar{\cal D}_j {\cal E}_{ik})  -  (\bar{\nabla}_j {\cal D}_i d   + \nabla _i \bar{\cal D}_j d) \\ [2.0ex] 
& &+ \ \ftb{1}{4} g^{nk} (\bar{\cal D}_j {\cal E}_{in} {\cal D}^m {\cal E}_{mk} + \ftb{1}{2} \bar{\cal D}_j {\cal E}_{lk} {\cal D}^l {\cal E}_{in} + {\cal D}_i {\cal E}_{nj} \bar{\cal D}^l {\cal E}_{kl}  + \ftb{1}{2} {\cal D}_i {\cal E}_{kl} \bar{\cal D}^l {\cal E}_{nj})  \\ [2.0ex]
& &- \ \ftb{1}{4}  (\bar{\cal D}^k {\cal E}_{lj} {\cal D}^l {\cal E}_{ik} + \ftb{1}{2} \bar{\cal D}^l {\cal E}_{kl} {\cal D}^k {\cal E}_{ij} + \ftb{1}{2} \bar{\cal D}^l {\cal E}_{ij} {\cal D}^k {\cal E}_{kl})  - \ftb{1}{4} g^{nk} g^{ml} \bar{\cal D}_j {\cal E}_{nm} {\cal D}_i {\cal E}_{kl}=0 .
\ea
\eq

 Here the $O(D, D)$ covariant derivatives $\nabla(\Gamma)$ and $\bar\nabla(\Gamma)$ are defined as follows \cite{Hohm-Hull-Z}:
 \bq
 \label{cov-der-gamma}
\ba{lcl}
\nabla_i(\Gamma) \bar{A}_j & \equiv & {\cal D}_i \bar{A}_j - \Gamma_{i \bar{j}}^{\bar{k}} \, \bar{A}_k \, , \\ [1.0ex]
\bar{\nabla}_j(\Gamma) {A}_i & \equiv & \bar{\cal D}_j {A}_i - \Gamma_{ \bar{j} i }^{k} \, {A}_k \, ,
\ea
\eq
where $A,\bar A$ are arbitrary vectors and the Christoffel-like symbols are defined in Appendix A.
The notation of barred and unbarred indices is related with the frame-like formalism and explained in the appendix A. \\

In the quest for a better understanding of these equations of motion and its possible geometric formulation the authors of \cite{Hohm-Hull-Z-1} proposed  a new formulation of the action in terms of a generalized metric ${\cal H}_{NM}$ that is not linear in the fields $g,b$. Surprisingly  using this formulation the symmetries of the theory are now linearly 
realized. A different covariant approach to compute  the double field equations of motion (\ref{eqm-R}) and (\ref{eqm-K}) using  $O(D,D)$ covariant indices leads to a form that resembles the Einstein field equations  \cite{korea} .

\section{Generalized Bianchi identities and differential identities}

The main result of our note is the construction of the generalized differential identities. We star from the  generalized Bianchi identities. In Appendix A we present a clear deduction of this identities using the frame-like formalism of DFT based on \cite{Hohm-Kwak}. These identities can be written as
\be\label{Bianchi4.1}
     \nabla_{a}{\cal R}+\nabla^{\bar{b}}{\cal R}_{a\bar{b}} \ = \ 0\;, \qquad
     \nabla_{\bar{a}}{\cal R}-\nabla^{b}{\cal R}_{b\bar{a}} \ = \ 0\;,
 \ee
in terms of `flat indices' $a,\bar a,b,\bar b$ (see Appendix A for details).  Using the vielbeins and fixing a particuler gauge to identify flat with curved indices, we can rewrite these identities in terms of curved indicies.
The Bianchi identities (\ref{Bianchi4.1}) in terms of curved indices are \cite{Kwak}
 \be\label{Bianchi-ij}
  \nabla^{i}{\cal R}_{ij}+\frac{1}{2}\bar{\cal D}_{j}{\cal R}({\cal E},d) \ = \ 0\;, \qquad
  \bar{\nabla}^{j}{\cal R}_{ij}+\frac{1}{2}{\cal D}_{i}{\cal R}({\cal E},d) \ = \ 0\;,
 \ee
where the covariant derivative is defined as
\be
\nabla_i=\nabla_i(\Gamma) - \ftb{1}{2} (\bar{\cal D}^k {\cal E}_{ik} + 4 {\cal D}_i d ),\quad
\bar\nabla_i=\bar{\nabla}_i(\Gamma) - \ftb{1}{2} ({\cal D}^k {\cal E}_{ki} + 4 \bar{\cal D}_i d ), 
\ee
and 
$\nabla_i(\Gamma)$, $\bar\nabla_i(\Gamma)$  covariant derivatives defined in (\ref{cov-der-gamma}) when acting on vectors.
The generalized Bianchi identities can be written then as (see also \cite{Kwak,Blair})
\be
\label{gen-bianchi-1}
\bigg[ {\nabla}^i (\Gamma)- \ftb{1}{2} g^{ik} (\bar{\cal D}^l {\cal E}_{kl} + 4 {\cal D}_k d) \bigg] {\cal K}_{ij} + \ftb{1}{2} \bar{\cal D}_j {\cal R} &=& 0 \label{final_bi1} \, , \\
\bigg[ \bar{\nabla}^j(\Gamma) - \ftb{1}{2} g^{jk} ({\cal D}^l {\cal E}_{lk} + 4 \bar{\cal D}_k d) \bigg] {\cal K}_{ij} + \ftb{1}{2} {\cal D}_i {\cal R} &=& 0  \label{gen-bianchi-2} \, .
\ee
We want to  stress that these expressions coincide with the Noether identities associated with the transformations
\be 
\delta {\cal E}_{ij} = \nabla_i(\Gamma) \bar{\eta}_j + \bar{\nabla}_j(\Gamma) \eta_i    \, , \quad
\delta d = - \ftb{1}{4}  \nabla_i  \eta^i 
-  \ftb{1}{4} \bar{\nabla}_i   \bar{\eta}^i  \, .
\ee
Apparently these symmetry transformations are not the original ones (\ref{gauge-transf}). But using a redefinition of the gauge parameters given by $\eta_i \equiv - \tilde{\xi}_i + {\cal E}_{ij} \xi^j$ and $\bar{\eta}_i \equiv \tilde{\xi}_i + {\cal E}_{ji} \xi^j$,  it is easy to see that these symmetries coincide with the original transformations (\ref{gauge-transf}).

 Following our construction of the subsection 2.1 we can define the  $\beta$ functions of DFT as
 \bq
 \label{ansatz}
{\cal K}_{ij} =a\beta_{ij} , \qquad   {\cal R}=b\beta^d,
 \eq
 for some constants $a,b$ that will be fixed later by requiring that the formalism coincide with the standard formalism when we turn off the $\tilde x$ variables. We will show in the next subsection that the coefficients are $a=-1$, $b=-4$, so the generalized differential identities are
 \be
 \label{gen-diff-iden}
{\nabla}^i  \beta_{ij} + 2 \bar{\cal D}_j {\beta^d} = 0  \, , \quad
 \bar{\nabla}^j  {\beta}_{ij} +{2} {\cal D}_i {\beta^d} = 0   \, .
\ee
These relations should be compared with the integrability conditions worked out in \cite{Berman}. 
As we will see in the next subsection these differential identities reduce to the differential identities (\ref{ident-dif-gauge-d}) and (\ref{ident-dif-difeos-d}) when we turn off the $\tilde x$ dependence. The same is true for the Bianchi identities and the equations of motion  ${\cal R}=0$ and ${\cal K}_{ij}=0$. With no $\tilde x$ dependence they reduce to the Bianchi identities and equations of motion of the effective action respectively.

\subsection{Turning off the $\tilde{x}$ dependence}

A remarkable statement of DFT is that we can recover the original effective action when we turn off the $\tilde x$ dependence by setting $\tilde\partial=0$. This case is just a particular frame in $O(D,D)$ (see footnote 2). This statement is also a consistency condition. In this section we will use this consistency condition to show that the proposed generalized differential identities of DFT (\ref{gen-diff-iden}) goes back to the original differential identities (\ref{ident-dif-gauge-d}, \ref{ident-dif-gauge-d}). As an important consequence of this condition, we will fix the coefficients $a$ and $b$ in (\ref{ansatz}) 

First we notice that the calligraphic derivatives defined in \eqref{cali-D} simplify as
\bq
{\cal D}_i = \bar{\cal D}_i = \partial_i \, .
\eq
Starting from the Bianchi identity (\ref{gen-bianchi-1}) (the other identity can be worked out in the same lines)
\begin{equation}
\bigg[\nabla^i(\Gamma)-\frac{1}{2}g^{il}(\bar{\mathcal{D}}^k\mathcal{E}_{lk}+4\mathcal{D}_ld)\bigg]\mathcal{K}_{ij}+\frac{1}{2}\mathcal{D}_j\mathcal{R}=0,
\label{Identidad-bianchi-wak}
\end{equation}
and using the definition of the covariant derivative (\ref{cov-der-gamma}) with
\begin{equation}\label{christoffel-main}
\Gamma^{\bar{k}}_{i\bar{j}}=\frac{1}{2}g^{kl}(\mathcal{D}_i\mathcal{E}_{lj}+\bar{\mathcal{D}}_j\mathcal{E}_{il}-\bar{\mathcal{D}}_l\mathcal{E}_{ij}), \quad
\Gamma^k_{ij}=\frac{1}{2}g^{kl}\mathcal{D}_i\mathcal{E}_{jl},
\end{equation}
the generalized Bianchi identity (\ref{Identidad-bianchi-wak}) can be written as,
\begin{equation}
\begin{split}
g^{in}&\bigg[\mathcal{D}_n\mathcal{K}_{ij}-\frac{1}{2}g^{lk}\mathcal{D}_n\mathcal{E}_{ik}\mathcal{K}_{lj}-\frac{1}{2}(\mathcal{D}_n\mathcal{E}_{kj}+\bar{\mathcal{D}}_j\mathcal{E}_{nk}-\bar{\mathcal{D}}_k\mathcal{E}_{nj})\mathcal{K}_{il}
-\frac{1}{2}(\bar{\mathcal{D}}^r\mathcal{E}_{nr}+4\mathcal{D}_n d)\mathcal{K}_{ij}\bigg]+\frac{1}{2}\bar{\mathcal{D}}_j\mathcal{R}=0.
\end{split}
\end{equation}

By turning off the $\tilde x$ dependence,  recalling that \(\mathcal{E}_{ij}=g_{ij}+b_{ij}\) and  $  {\cal K}_{ij}=a\beta_{ij}$,  $ {\cal R}=b\beta^d$,
we find
\begin{equation}
a\nabla^i\beta^g_{ij}-a2\nabla^i d\beta^g_{ij}-\frac{a}{2}H_{jli}\beta^{b\ li}+\frac{b}{2}\partial_j\beta^d+a\nabla^i\beta^{b}_{ij}-a2\nabla^i d\beta^b_{ij}=0,
\end{equation}
where $\beta^g_{ij}$ and $\beta^b_{ij}$ are the symmetric and the antisymmetric part of $\beta_{ij}$, when $\tilde \partial$ is set to zero. To recover from here the original differential identities (\ref{ident-dif-gauge-d}, \ref{ident-dif-difeos-d}), we need to fix $a=-1$, $b=-4$.

Our result is consistent with the previously reported results about the reduction of the equations of motion of DFT when $\tilde\partial=0$. Indeed
when we turn off $\tilde x$ the $O(D,D)$ $R$ scalar is given in~\cite{Hohm-Hull-Z}. The explicit expression is
\bq
{\cal R} \Big|_{\tilde{\partial}=0} = R + 4(\nabla^i \nabla_i d - (\nabla d)^2) - \ftb{1}{12}H^2 \, .
\eq
Thus the equation of motion of the dilaton $d$ in double field theory with no $\tilde{x}$ dependence is
\bq
 \label{beta-d-1}
\beta^d=-4(R + 4(\nabla^i \nabla_i d - (\nabla d)^2) - \ftb{1}{12}H^2) \, ,
\eq
that coincide exactly with the equation of motion derived in our previous section for the dilaton $d$. This result imply $b=-1/4$ in agreement with our result.

The other equations of motion ${\cal K}_{ij}$  reduce to \cite{Kwak}
\bq
\label{K-beta}
\ba{l} 
{\cal K}_{ij} \Big|_{\tilde{\partial}=0} = -\Bigg[  R_{ij}  - \ftb{1}{4} {H_{i}}^{k l} H_{j kl}  + 2 \nabla_i \nabla_j d \Bigg] - \Bigg[ -\ftb{1}{2} \nabla^k H_{kij} + H_{kij} \nabla^k d \Bigg] \, .
\ea
\eq
Notice that the terms in the first bracket are symmetric under while the terms in the second bracket are antisymmetric. 

From (\ref{K-beta}) we recognize the $\beta$ functions of the effective action in terms of the new dialton $d$ 
\bq
{\cal K}_{ij} \Big|_{\tilde{\partial}=0} = -\beta^G_{ij} - \beta^B_{ij} \, .
\eq
This result imply $a=-1$ in agreement with our result.

It is also true that when we turn off the $\tilde{x}$-dependence the generalized Bianchi identities reduce to \cite{Kwak}
\bq \label{Bianchi}
 \ftb{1}{2} \bigg[ \nabla^i \nabla^n H_{nij}  + \Big( \ftb{1}{2}  {H_i}^{l n} \nabla^i H_{j l n} - \ftb{1}{12} \nabla_j H^2 \Big)  \bigg] \\ + \bigg( -\nabla^i R_{ij} + \ftb{1}{2} \nabla_j R \bigg) =0 \, ,
\eq
\bq \label{Bianchi2}
 \ftb{1}{2} \bigg[ \nabla^j \nabla^n H_{nij}  + \Big( \ftb{1}{2}  {H_j}^{ l n} \nabla^j H_{i  l n} - \ftb{1}{12} \nabla_i H^2 \Big)  \bigg] \\ + \bigg( -\nabla^j R_{ij} + \ftb{1}{2} \nabla_i R \bigg) =0 \, .
\eq
From here also we can recognize the Bianchi identities (\ref{gauge-bianchi}, \ref{GR-bianchi}, \ref{H-bianchi}) and  the differential identities.
The Noether identities (\ref{ident-noether-gauge-d}, \ref{ident-noether-difeos-d}) also follow from here by the identification of the equations of motion with the $\beta$ functions. Finally we can reconstruct the gauge and space-time symmetries (\ref{sym-g-d}, \ref{sym-b-d}, \ref{sym-d}) of the effective action (\ref{action-d}).

Our results confirms that the equations of motion of DFT are the generalized $\beta$ functions as was reported from the results of the double sigma model in \cite{Berman1, sigma-model}.

\section{Conclusions}

We worked out a relation between $\beta$ functions of the closed string sigma model and the Noether identities. The $\beta$ functions have an interesting property: they have encoded information about the symmetries of the effective action. Indeed through the differential identities and the $\beta$ function relations with equations of motion of the effective action, the Noether identities can be constructed and from them the symmetries of the effective action can be retrieved. The Noether identities can also be identified with the Bianchi and the differential identities.
We develop, as an application of these ideas the recovering of the symmetries of the DFT action. The structure of the symmetries is highly non trivial but nevertheless the powerful of the Noether second theorem can be applied. As a consequence of our results, we found the generalized differential identities in DFT. As a check for consistency of our results we have contrasted them  with previous works.

The new dilaton $d$ plays a crucial role in our formulation. 
Another interesting area where these ideas can be explored is in the study of the dynamics of a particle in the metric background given by ${\cal H}_{MN}$. Using this metric the authors of \cite{geo-H} constructed a generalized geodesic equation. We want to explore the symmetries and application of such geodesic flux in the near future. Another interesting line that we want to explore is the first order formulation of the DFT and the duality relations that can be constructed at the linealized level \cite{Bergshoeff}. In this context we want to explore the relation between DFT action and the partially massless background. Finally we left for a future work to check the consistency of the gauge fixing in the frame-like formalism to relate the first order formulation with alternative formulations of DFT. In particular the ADM formulation of the DFT could help to understand this issue \cite{Naseer}.

\section{Aknowledgements}

This work  was partially supported by Mexico National Council of Science and Technology (CONACyT) grant CB 2015-238734-F. RA was partially supported by a fellowship number 744575 CONACyT.

\section{Appendix A}

We will review DFT using the frame like formalism. A very  clear exposition of these ideas can be found in \cite{Hohm-Kwak} where we refer the reader for the details. All our notation and convetions are taken from \cite{Hohm-Kwak}. Here we only give a very brief review of the basic concepts that are relevant for our purpose. In particular we want to explain the concept of integration by parts in DFT and the construction of the Bianchi identities.

Using frame-like notation the $O(D,D)$ invariant metric $\eta_{MN}$ induces
an $X$-dependent tangent space metric given by
 \be
  {\cal G}_{AB} \ = \ e_{A}{}^{M}\,e_{B}{}^{N}\,\eta_{MN}\;,
 \ee
with inverse ${\cal G}^{AB}=\eta^{MN}e_{M}{}^{A}e_{N}{}^{B}$. We have introduced a frame field $e_{A}{}^{M}$,
with flat index $A$ corresponding to a local
$GL(D)\times GL(D)$ symmetry, 
 \be\label{Egauge}
   e_{A}{}^{M} \ = \ \begin{pmatrix} e_{ai} &  e_{a}{}^{i} \\ e_{\bar{a}i} & e_{\bar{a}}{}^{i} \end{pmatrix}
   \;,
 \ee
and assume that this vielbein is invertible, denoting the inverse by $e_{M}{}^{A}$
\footnote{Notation: The index $M = (\,{}_{i}\,,\,{}^{i}\,)$ is of the $O(D,D)$. The index  $A=(a,\bar{a})$
is an $GL(D)\times GL(D)$ index.}. 

Interestingly enough the $O(D,D)$ is a global symmetry that maps solutions into solutions of the DFT action by relating different backgrounds $g, b,d\to g',b',d'$ as the Buscher rules. In turn each 'point' in this solution space is enhanced with gauge and space time local symmetries.

Using the frame field one can introduce a derivative $e_{A}$, defined by 
 \be\label{flatder}
  e_{A} \ \equiv \ e_{A}{}^{M}\,\partial_{M}\;.
\ee
Now we can introduce covariant
derivatives 
 \be\label{covder}
  \nabla_{A}V_B \ = \ e_{A}V_{B}+\omega_{AB}{}^{C}V_{C}\;, \qquad
  \nabla_{A}V^{B} \ = \ e_{A}V^{B}-\omega_{AC}{}^{B}V^{C}\;,
 \ee
where we have introduced connections $\omega_{AB}{}^{C}$.

The covariant derivative is defined by a torsion free, and compatibility condition $
  \nabla_{A}{\cal G}_{BC} \ = \ 0 .$
We also require the condition of integration by parts 
 \be\label{dilconstr}
  \int e^{-2d}\, V\nabla_{A}V^{A} \ = \ -\int e^{-2d}\,V^{A}\nabla_{A}V \ = \
  -\int e^{-2d}\,V^{A}e_{A}V\; ,
 \ee
for arbitrary $V$ and $V^{A}$. From this last condition we find
 \be\label{tracepart}
  \omega_{BA}{}^{B} \ = \ -e^{2d}\partial_{M}\big(e_{A}{}^{M}e^{-2d}\big)
  \ = \ -\partial_{M}e_{A}{}^{M}+2e_{A}d\;,
 \ee

To construct the Bianchi identities the starting point is the  action 
\be\label{actionfinal}
   S \ = \ \int dxd\tilde{x}\,e^{-2d}\,{\cal R}\;.
 \ee
 Introducing a flat variation
 \be
 \label{DeltaAB}
 \Delta_{AB}=e_B{}^M\delta e_{AM}.
 \ee
 To describe the correct physical degrees of freedom using as dynamical variables the frames $e_A{}^M$,  it turns out necessary to impose the constraint
 \be
 {\cal G}_{a\bar b}=0,
 \ee
 or $e_a{}^i e_{\bar b}{}^i+e_{\bar b}{}^i e_{ a}{}^i=0$.
The variation of (\ref{actionfinal})  associated to the variation on the frames can be written as
 \be\label{GenVar}
   \delta S \ = \ \int dxd\tilde{x}\,e^{-2d}\left(-2\delta d\,{\cal R}+
   \Delta e_{a\bar{b}}{\cal R}^{a\bar{b}}\right)\;,
 \ee
and the field equations are then
 \be
   {\cal R} \ = \ 0\;, \qquad {\cal R}_{a\bar{b}} \ = \ 0\;.
 \ee
The symmetries of the action in terms of the frames are,
 \be\label{deltacov}
   \delta_{\xi}e_{A}{}^{M} \ = \ -e_{B}{}^{M}\left(\nabla_{A}\xi^{B}-\nabla^{B}\xi_{A}\right)
      \;.
 \ee
The gauge symmetry for the frames is then
\be
\Delta_{AB}=\nabla_B\xi_A+\nabla_A\xi_B.
\ee
The relevant components that are used in the variation of the action are
\be \label{symm-frame}
\Delta_{a\bar b}=\nabla_{\bar b}\xi_a+\nabla_a\xi_{\bar b}.
\ee
The dilaton symmetry transformations given in the main text (\ref{gauge-transf}) can be written in terms of frames as
 \be\label{covd}
  \delta_{\xi}d \ = \ -\frac{1}{2}\nabla_{A}\xi^{A}\ = \  -\frac{1}{2}\left(e_{A}\xi^{A}-\omega_{BA}{}^{B}\xi^{A}\right).
  \ee
The Bianchi identity that comes from the gauge invariance of the action (\ref{actionfinal}) can be constructed as follows  \cite{Blair}. First notice that
 \be
 \begin{split}\label{var-symm}
   \delta_{\xi}S \ &= \ \int dx d\tilde{x}\,e^{-2d}\left(\left(\nabla_{a}\xi^{a}+\nabla_{\bar{a}}\xi^{\bar{a}}\right)
  {\cal R}+\left(\nabla_{\bar{b}}\xi_{a}-\nabla_{a}\xi_{\bar{b}}\right){\cal R}^{a\bar{b}}\right) \\
  \ &= \ -\int dxd\tilde{x}\,e^{-2d}\left(\xi^{a}\left(\nabla_{a}{\cal R}+\nabla^{\bar{b}}{\cal R}_{a\bar{b}}\right)
  +\xi^{\bar{a}}\left(\nabla_{\bar{a}}{\cal R}-\nabla^{b}{\cal R}_{b\bar{a}}\right)\right)\;,
 \end{split}
 \ee
where we used the variation of the action (\ref{actionfinal}) specialized for a variation of the gauge symmetry (\ref{symm-frame},\ref{covd}) with parameters $\xi^a$ and integrate by parts in the second line. As the action is gauge invariant $\delta_\xi S=0$, (\ref{var-symm}) imply \cite{Siegel}
 \be\label{Bianchi4}
     \nabla_{a}{\cal R}+\nabla^{\bar{b}}{\cal R}_{a\bar{b}} \ = \ 0\;, \qquad
     \nabla_{\bar{a}}{\cal R}-\nabla^{b}{\cal R}_{b\bar{a}} \ = \ 0\;.
 \ee
 To see that these Bianchi identities are in fact the Bianchi identites given in the main text (\ref{gen-bianchi-1}, \ref{gen-bianchi-2}) we need to relate this frame-like formalism with the ${\cal E}$ formalism. For that end we first observe that the gauge choice
 \be\label{Egauge2}
   e_{A}{}^{M} \ = \ \begin{pmatrix} e_{ai} &  e_{a}{}^{i} \\ e_{\bar{a}i} & e_{\bar{a}}{}^{i} \end{pmatrix}
   \ = \ \begin{pmatrix} -{\cal E}_{ai} &   \delta_{a}{}^{i} \\ {\cal E}_{i\bar{a}} & \delta_{\bar{a}}{}^{i} \end{pmatrix}\;,
 \ee
allow the identification of the space-time indices $i,j,\ldots$  with the frame-like indices of either
$GL(D)$. These vielbeins $e_a^{\ i}=\delta_{a}{}^{i}$ or $e_{\bar a}^{\ i}=\delta_{\bar{a}}{}^{i}$ implement such identification. The derivatives (\ref{cali-D}) then coincide with the partial derivatives (\ref{flatder}),
\be\label{callderflat}
  {\cal D}_{a} =e_{a} \ = \ e_{a}{}^{M}\partial_{M} \ = \ \partial_{a}-{\cal E}_{ai}\tilde{\partial}^{i} \;, \qquad
  \bar{\cal D}_{\bar{a}}=e_{\bar{a}} \ = \ e_{\bar{a}}{}^{M}\partial_{M} \ = \ \partial_{\bar{a}}+{\cal E}_{i\bar{a}}\tilde{\partial}^{i} \;.
 \ee
 We have two identifications for the metric $g_{ij}={\cal E}_{(ij)}$ in  the tangent space $GL(D)\times GL(D)$ acording to either sector of $GL(D)$
 \be\label{gs}
  g_{ab} \ = \ -\frac{1}{2}e_{a}{}^{M}e_{b}{}^{N}\eta_{MN}\;, \qquad
  g_{\bar{a}\bar{b}} \ = \ \frac{1}{2}e_{\bar{a}}{}^{M}e_{\bar{b}}{}^{N}\eta_{MN}\;.
 \ee
Now  ${\cal G}_{AB}$ is given by
 \be
  {\cal G}_{AB} \ = \  \begin{pmatrix} -2g_{ab} &  0 \\ 0 & 2g_{\bar{a}\bar{b}} \end{pmatrix}\;.
 \ee
Recalling the redefinitions of the gauge parametres as
 \be\label{SFTbasis}
   \eta_{i} \ = \ -\tilde{\xi}_{i}+{\cal E}_{ij}\xi^{j}\;, \qquad
   \bar{\eta}_{i} \ = \ \tilde{\xi}_{i}+\xi^{j}{\cal E}_{ji}\;,
 \ee
and the gauge transformations  (\ref{gauge-transf}) given in the main text
 \be\label{delcalE}
  \delta{\cal E}_{ij} \ = \ \nabla_{i}(\Gamma)\bar{\eta}_{j}+\bar{\nabla}_{j}(\Gamma)\eta_{i}\;,
\ee
where $\nabla_{i}(\Gamma)$ are the covariant derivatives defined through the Christoffel-like connexions (\ref{christoffel-main}).
The relations with the $GL(D)\times GL(D)$
connections is
 \be\label{etabase}
  \eta_{a} \ := \ -\xi_{a} \ \equiv \ -e_{a}{}^{M}\xi_{M}\;, \qquad
  \bar{\eta}_{\bar{a}} \ := \ \xi_{\bar{a}} \ \equiv \ e_{\bar{a}}{}^{M}\xi_{M}\;,
 \ee
which coincide with (\ref{SFTbasis}) after the gauge fixing  (\ref{Egauge2}). The gauge symmetries for ${\cal E}$ are 
 \be\label{gaugeA}
  \delta  {\cal E}_{a\bar{b}} 
  \ = \  \nabla_{a}\bar{\eta}_{\bar{b}}+\nabla_{\bar{b}}\eta_{a}\;.
 \ee
Using the frame formalism we can check that the original form  of the gauge transformations (\ref{gauge-transf}) can be recovered from (\ref{gaugeA}).
  
The $O(D,D)$ covariant derivatives coincide with the
$GL(D)\times GL(D)$ covariant derivatives after gauge fixing. 
The Christofel-like symbols are
  \be
    \Gamma_{i \bar{j}}^{\bar{k}}  = \frac{1}{2} g^{kl} \Big( {\cal D}_i {\cal E}_{lj}
    + \bar{\cal D}_j {\cal E}_{il} - \bar{\cal D}_l {\cal E}_{ij}  \Big)  , \quad
     \Gamma_{\bar{i} j}^{k}  = \frac{1}{2} g^{kl} \Big( \bar{\cal D}_i {\cal E}_{jl}
     + {\cal D}_j {\cal E}_{l i} - {\cal D}_l {\cal E}_{ji}  \Big) ,
    \ee
    and the traces are given by
    \be
      \Gamma_{ji}^{j}= \frac{1}{2} {\cal D}^{j} {\cal E}_{ij}, \quad
  \Gamma_{\bar{j} \bar{i}}^{\bar{j}}=\frac12 \bar{\cal D}^{j} {\cal E}_{ji}.
\ee
 The connexions that can be determined from the Christofel-like symbols is 
     \begin{eqnarray}
     \omega_{i \bar{j}}{}^{\bar{k}}  &=&  - \Gamma_{i \bar{j}}^{\bar{k}} \,,\\ 
 \omega_{\bar{i} j}{}^{k}  &=&    - \Gamma_{\bar{i} j}^{k},\\
  \omega_{ji}{}^{j} &=& - \Gamma_{ji}^{j} + \frac{1}{2} \bar{\cal D}^{j}{\cal E}_{ij} + 2 {\cal D}_i d \, ,
      \\
      \omega_{\bar{j} \bar{i}}{}^{\bar{j}} &=& 
      - \Gamma_{\bar{j} \bar{i}}^{\bar{j}} + \frac{1}{2}  {\cal D}^{j} {\cal E}_{ji} + 2 \bar{\cal D}_i d   \, 
 .\end{eqnarray}
In this notation we obtain a simple gauge transformation of $d$ 
 \be\label{deldcov}
  \delta d \ = \ -\frac{1}{4}\nabla_{i}\eta^{i}-\frac{1}{4}\bar{\nabla}_{i}\bar{\eta}^{i}\;.
 \ee
Finally the definition of the covariant derivative in terms of the usual covariant derivative $\nabla(\Gamma)$ is
\be
\nabla_i= \nabla_i(\Gamma) - \ftb{1}{2} (\bar{\cal D}^k {\cal E}_{ik} + 4 {\cal D}_i d ),
\ee
and the Bianchi identities (\ref{Bianchi4}) can be written as
 \be
  \nabla^{i}{\cal R}_{ij}+\frac{1}{2}\bar{\cal D}_{j}{\cal R}({\cal E},d) \ =\  0\;, \qquad
  \bar{\nabla}^{j}{\cal R}_{ij}+\frac{1}{2}{\cal D}_{i}{\cal R}({\cal E},d) \ = \ 0\;,
 \ee
which agree with the Bianchi identities given in the main text (\ref{Bianchi-ij}).

\end{document}